\documentclass[conference]{IEEEtran}
\IEEEoverridecommandlockouts
\usepackage{cite}
\usepackage{amsmath,amssymb,amsfonts}
\usepackage{booktabs}
\usepackage{algorithmic}
\usepackage{graphicx}
\usepackage{subcaption}
\usepackage{svg}
\usepackage{textcomp}
\usepackage{xcolor}
\usepackage{soul}
\usepackage{rotating}
\usepackage{stfloats}
\usepackage[skip=0pt]{caption}

\usepackage{physics}
\usepackage{comment}
\usepackage{orcidlink}
\def\BibTeX{{\rm B\kern-.05em{\sc i\kern-.025em b}\kern-.08em
    T\kern-.1667em\lower.7ex\hbox{E}\kern-.125emX}}

\begin{document}

\title{Benchmarking VQE Configurations: Architectures, Initializations, and Optimizers for Silicon Ground State Energy}

\author{\IEEEauthorblockN{Z. Boutakka\textsuperscript{1}, N. Innan\orcidlink{0000-0002-1014-3457}\textsuperscript{2,3},  M. Shafique\orcidlink{0000-0002-2607-8135}\textsuperscript{2,3}, M. Bennai\orcidlink{0000-0002-7364-5171}\textsuperscript{1}, Z. Sakhi\textsuperscript{1}
}
\IEEEauthorblockA{\textsuperscript{1}Quantum Physics and Spintronics Team, LPMC, Faculty of Sciences Ben M'sick,\\ Hassan II University of Casablanca,
Morocco\\
\textsuperscript{2}eBRAIN Lab, Division of Engineering, New York University Abu Dhabi (NYUAD), Abu Dhabi, UAE\\
\textsuperscript{3}Center for Quantum and Topological Systems (CQTS), NYUAD Research Institute, NYUAD, Abu Dhabi, UAE\\
zakaria.boutakka-etu@etu.univh2c.ma,  nouhaila.innan@nyu.edu, \\
muhammad.shafique@nyu.edu,  mohamed.bennai@univh2c.ma, zb.sakhi@gmail.com\\
}}

\maketitle

\begin{abstract}
Quantum computing presents a promising path toward precise quantum chemical simulations, particularly for systems that challenge classical methods. This work investigates the performance of the Variational Quantum Eigensolver (VQE) in estimating the ground-state energy of the silicon atom, a relatively heavy element that poses significant computational complexity. Within a hybrid quantum-classical optimization framework, we implement VQE using a range of ansatz, including Double Excitation Gates, ParticleConservingU2, UCCSD, and k-UpCCGSD, combined with various optimizers such as gradient descent, SPSA, and ADAM. The main contribution of this work lies in a systematic methodological exploration of how these configuration choices interact to influence VQE performance, establishing a structured benchmark for selecting optimal settings in quantum chemical simulations. Key findings show that parameter initialization plays a decisive role in the algorithm’s stability, and that the combination of a chemically inspired ansatz with adaptive optimization yields superior convergence and precision compared to conventional approaches.
\end{abstract}


\section{Introduction}

In recent years, Artificial Intelligence (AI) has dominated technological discourse. However, increasing attention is now being directed toward Quantum Computing (QC) as an emerging field that promises to revolutionize computation by exploiting the principles of quantum mechanics \cite{Vogel}. With advancements in qubit design and quantum hardware \cite{google2025quantum, microsoft2025interferometric}, the practical realization of QC is becoming more tangible.
Among the most promising applications of QC is quantum chemistry \cite{Aspuru}, which aims to predict and explain the properties of matter at atomic and molecular scales. Atoms, the fundamental constituents of all compounds, combine to form molecules, , whose interactions determine the behavior of nature and the complexity of materials. Determining electronic structures allows researchers to predict system behavior and material properties. These calculations are based on quantum mechanics, particularly the time-independent Schrödinger equation. Because this equation is a high-dimensional and generally nonlinear partial differential equation, it rarely has closed-form solutions for realistic systems and is typically solved using approximate computational algorithms.

In this context, the Hamiltonian describes the interactions between electrons and nuclei. The wavefunction, which is the solution to the Schrödinger equation, contains all measurable information about a quantum system. The ground-state energy, representing the lowest energy configuration of the system, is of central importance for understanding molecular behavior, designing new materials, and optimizing chemical reactions.
Traditionally, these problems have been addressed using classical computational methods such as Hartree–Fock (HF) \cite{Lewars, Szabo, Jensen}, Density Functional Theory (DFT) \cite{Christopher, Yu, Mardirossian}, and Full Configuration Interaction (FCI) \cite{Dobrautz}. HF is computationally efficient but does not fully capture electron correlation effects \cite{Belaloui}. FCI provides exact solutions within a chosen basis set but scales exponentially with system size, making it impractical for larger systems. DFT offers a compromise between precision and efficiency, but it still relies on approximations and can struggle with scaling and precision in strongly correlated systems \cite{Pavlo, Gao, Troyer}. These limitations highlight the need for more efficient and precise approaches.

To overcome these limitations, Peruzzo et al. \cite{Peruzzo} introduced the Variational Quantum Eigensolver (VQE), a hybrid quantum–classical algorithm designed to estimate the ground-state energy of a quantum system. VQE is particularly well suited for Noisy Intermediate-Scale Quantum (NISQ) devices \cite{Preskill}, which are characterized by limited qubit counts and relatively high error rates. The approach employs parameterized quantum circuits (ansatz) to prepare trial wavefunctions and uses classical optimization algorithms to minimize the expectation value of the Hamiltonian.
VQE has become a promising alternative to classical methods for solving the electronic structure problem. By variationally optimizing the ansatz parameters, it can explore high-dimensional Hilbert spaces more efficiently. However, its performance strongly depends on the choice of ansatz architecture, classical optimization algorithm, and parameter initialization strategy \cite{Tilly,Pellow-Jarman,Sorourifar,Chen}.

Despite its potential, VQE faces several challenges. A major bottleneck is the cost of measuring Hamiltonian expectation values, which scales unfavorably for complex systems \cite{Belaloui}. In addition, the optimization process is nontrivial and, in many cases, NP-hard \cite{Bittel}. Problems such as barren plateaus, which are regions in parameter space where gradients vanish exponentially, can slow convergence \cite{Larocca}, particularly for deep circuits and large systems. Although mitigation strategies, such as identity block initialization or local Hamiltonian encoding, have been proposed, their effectiveness for medium- to large-scale systems remains an open question.

To address this, we focus on the silicon atom, an important system in this context, playing a key role in materials science and semiconductor technology. Accurately computing its ground-state energy is a computationally demanding task due to the system's complexity and the large Hilbert space it occupies. Classical methods can, in principle, determine this energy, indeed, high-precision techniques such as FCI \cite{Dobrautz} and Coupled Cluster [CCSD(T)] \cite{Bartlett} have estimated it at approximately -289 Ha \cite{NIST, CODATA}, but such approaches require enormous computational resources and scale exponentially with system size. More approximate methods, such as HF or DFT, offer efficiency but may fail to capture electron correlation with sufficient precision.
In this regard, the VQE offers an attractive pathway for accurately estimating the silicon atom’s ground-state energy without the exponential cost of classical methods.  As a hybrid quantum-classical algorithm, VQE leverages quantum circuits to represent complex, entangled wavefunctions in exponentially large Hilbert spaces, while classical optimizers iteratively minimize the system’s energy expectation value. This synergy enables VQE to efficiently approximate the ground state of complex atoms like silicon with reduced computational overhead compared to exact classical methods. Furthermore, its flexibility through the choice of informed parameter initialization, shallow or chemically inspired ansatz, and adaptive optimization algorithms allows VQE to capture electronic correlations and achieve stable convergence toward the true ground-state energy. These characteristics make VQE feasible for near-term quantum hardware and a methodologically robust approach for exploring medium-scale atomic systems.

\begin{figure*}[h]
\centering
\includegraphics[width=\linewidth]{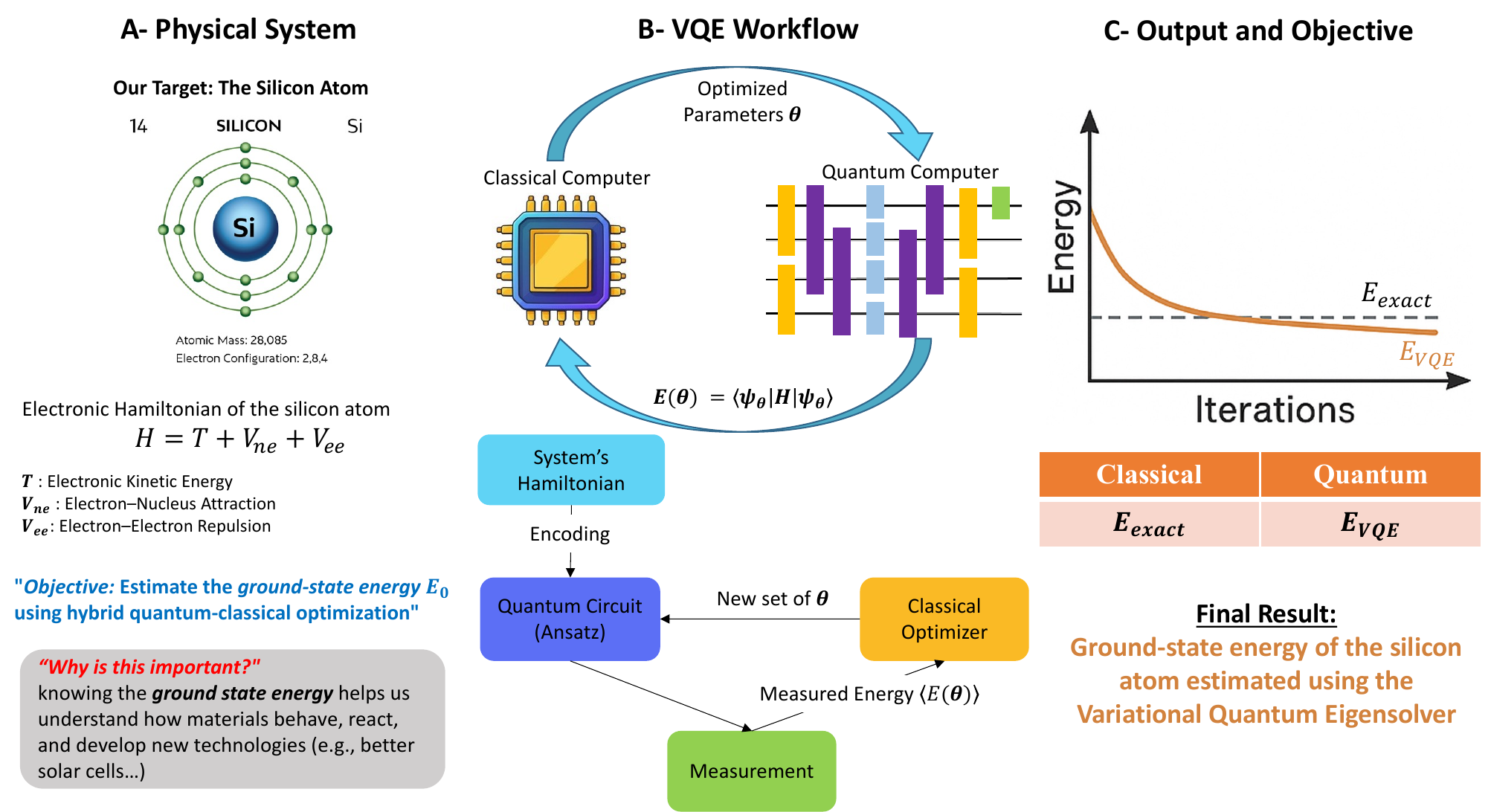}
\caption{Motivational overview of the VQE framework for estimating the ground-state energy of the silicon atom. Panel (A) presents the physical system and its electronic Hamiltonian, defining the objective of obtaining the ground-state energy $E_0$. Panel (B) illustrates the hybrid quantum--classical VQE loop used to approximate $E_0$ through parameterized quantum circuits and iterative optimization. Panel (C) shows the convergence of the estimated energy toward $E_{\text{exact}}$. While the ultimate goal is precise energy estimation, the effectiveness of VQE critically depends on several design choices, including ansatz structure, optimizer configuration, and parameter initialization, motivating the comparative analysis conducted in this study.}

\label{overview}
\end{figure*}

Consequently, VQE emerges as the most promising framework for studying the silicon atom and similar medium-scale systems, where classical approaches become computationally prohibitive. Yet, despite its potential, existing studies often lack a comprehensive comparative analysis of the factors that influence VQE performance \cite{Cerezo, Tilly, McClean}, particularly in terms of ansatz design, optimizer selection, and initialization strategies for medium-sized atoms such as silicon. As shown in Fig. \ref{overview}, the schematic provides a general overview of the system and problem under consideration, illustrating the physical model of the silicon atom, the hybrid quantum–classical VQE workflow used to approximate its ground-state energy, and the convergence behavior of the optimization process.

In this work, we address this gap by conducting a systematic investigation of the ground-state energy of the silicon atom using the VQE.
The main contributions of this study are:
\begin{itemize}
\item Examination of how different ansatz architectures affect VQE performance for a medium-sized atomic system.
\item Assessment of the combined influence of parameter initialization schemes and classical optimization algorithms.
\item Analysis of convergence behavior, ground-state energy estimates within chemical precision, and optimization stability.
\item Discussion of the relative importance of algorithmic design choices for practical hybrid quantum–classical computations.
\end{itemize}

Building on these contributions, our results demonstrate that parameter initialization plays a decisive role in determining VQE convergence behavior and final energy precision. Across all tested configurations, initializing parameters at zero consistently leads to faster and more stable convergence compared to other initialization strategies. Moreover, the combination of a chemically inspired ansatz (such as UCCSD) with adaptive optimization methods (notably ADAM) provides the most robust and precise ground-state energy estimations for the silicon atom. These outcomes highlight methodological improvements over current VQE implementations, emphasizing that careful configuration of parameter initialization, ansatz design, and optimization technique can substantially enhance both convergence and precision in quantum chemical simulations.

The rest of this paper is organized as follows. Section~\ref{sec2} reviews related work and establishes the theoretical and methodological foundation for our study. Section~\ref{sec3} describes the VQE architecture and outlines the methodology adopted. Section~\ref{sec4} presents the experimental setup and reports and analyzes the simulation results, comparing VQE performance with experimental data. Finally, Section~\ref{sec6} summarizes the key findings and discusses the broader implications of VQE for research in materials science and physics.

\section{Background and Related Work \label{sec2}}

\subsection{Background}

 VQE is a hybrid quantum–classical algorithm that has emerged as a practical approach for estimating the ground-state energies of molecular and material systems \cite{Innan, Sivakumar}. Based on the variational principle, VQE guarantees that the expectation value of the Hamiltonian for any trial wavefunction $\ket{\psi(\theta)}$ provides an upper bound to the true ground-state energy:
\begin{equation}
E(\theta) = \bra{\psi(\theta)} \hat{H} \ket{\psi(\theta)} \geq E_0,
\end{equation}
where $E(\theta)$ is the energy of the trial state $\ket{\psi(\theta)}$, $\hat{H}$ is the system Hamiltonian, $\theta$ is the set of circuit parameters, and $E_0$ is the exact ground-state energy.

The trial wavefunction is prepared by applying a parameterized quantum circuit (PQC), denoted $U(\theta)$, to an initial reference state $\ket{0}^{\otimes n}$:
\begin{equation}
\ket{\psi(\theta)} = U(\theta) \ket{0}^{\otimes n},
\end{equation}
where $n$ is the number of qubits, $\ket{0}^{\otimes n}$ is the $n$-qubit zero state, and $U(\theta)$ is composed of tunable quantum gates whose parameters are optimized to minimize $E(\theta)$.

\begin{figure*}[h]
\centering
\includegraphics[width=\linewidth]{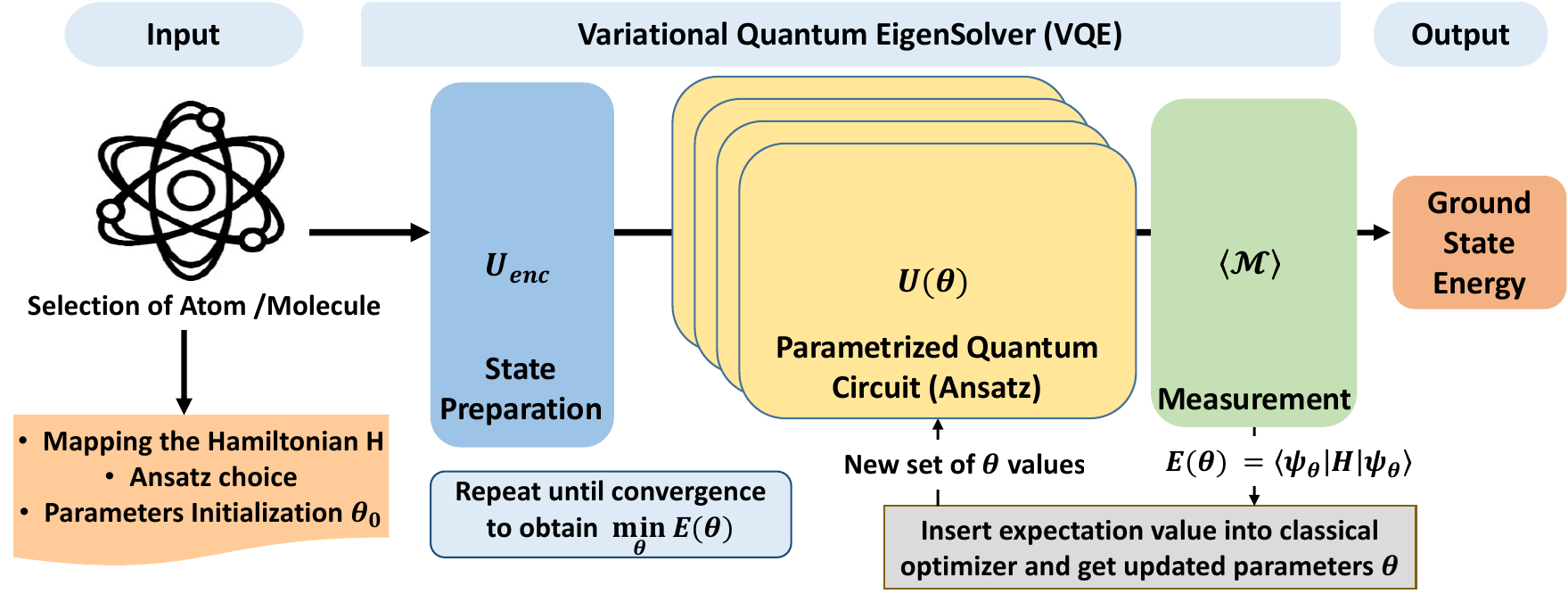}
\caption{Schematic of the VQE workflow, starting with the preparation of parameterized quantum states, then measuring observables and feeding the results to a classical optimizer, which iteratively updates the parameters.}
\label{vqe}
\end{figure*}

The optimization proceeds iteratively: at each step, the quantum processor prepares $\ket{\psi(\theta)}$, measures the expectation values of the Hamiltonian terms, and sends the results to a classical optimizer. The optimizer updates $\theta$ using methods such as gradient descent, stochastic gradient descent, or other heuristics until convergence (see Fig.~\ref{vqe}).

In quantum chemistry, the molecular electronic structure is described by the electronic Hamiltonian, which captures the kinetic energy of electrons, their attraction to nuclei, and electron–electron repulsion. To make this Hamiltonian amenable to quantum computation, it is typically expressed in a second-quantized form, where fermionic creation and annihilation operators represent the electronic degrees of freedom \cite{Magalhaes, McArdle}.

Since quantum computers operate on qubits rather than fermions, this Hamiltonian must be mapped into a qubit representation. Standard transformations such as Jordan–Wigner and Bravyi–Kitaev enable this conversion by expressing the fermionic operators in terms of tensor products of Pauli operators \cite{Jordan, Seeley}. The result is a qubit Hamiltonian written as a weighted sum of Pauli strings, which forms the cost function in VQE simulations.

The performance of VQE depends on the expressibility of the chosen ansatz, its ability to approximate the true ground state, and the robustness of the classical optimizer in navigating complex and possibly noisy cost landscapes. VQE has been applied successfully in quantum chemistry to approximate molecular ground-state energies, yielding insights into electronic structures and reaction dynamics. In materials science, it has been used to guide the design of materials with tailored properties for electronics, energy storage, and other applications \cite{Delgado, MaH, Mustafa}.

Building on these foundations, research continues to improve VQE through advances in ansatz design, optimization techniques, measurement reduction strategies, and the incorporation of Machine Learning (ML). The next section reviews these developments, tracing the progression of VQE from a theoretical construct to a practical tool for challenging electronic structure problems in molecular and material systems.

\subsection{Related Work}

Over the past few years, a considerable body of work has focused on the electronic structure problem in molecular systems, employing methods that range from foundational quantum algorithms to modern hybrid quantum–classical approaches. Among the earliest QC contributions,
Whitfield \textit{et al}. \cite{Whitfield} demonstrated the application of Quantum Phase Estimation (QPE) to determine the energy levels of small molecules such as hydrogen ($H_{2}$), and presented one of the foundational efforts in this direction. They accomplished this with precomputed molecular integrals but were limited by the assumption of nuclear configuration, more broadly highlighting the challenge of scalability in quantum simulations.

To overcome such limitations, attention has shifted in growing intensity to hybrid quantum-classical algorithms, particularly VQE. As a more NISQ-compatible method, VQE has opened the door for practical simulations on near-term quantum devices. Qing and Xie \cite{Qing} demonstrated the ability of VQE to accurately compute the ground state energy of the $H_{2}$ molecule using Qiskit and IBM Quantum hardware, while also noting the practical constraints associated with extending this method to more complex systems.

Parallel to these algorithmic developments, a wave of research began to explore the integration of ML with quantum computation to improve performance further and reduce resource demands. Carleo and Troyer \cite{Carleo} introduced a variational neural-network ansatz using perceptrons with hidden layers to represent quantum states in spin systems, employing reinforcement learning to optimize the model. Similarly, Xia and Kais \cite{Xia-Kais} proposed a hybrid framework that utilized a Restricted Boltzmann Machine (RBM) to approximate the ground state energies of small molecules. Their results, validated on molecules such as $H_{2}$, $LiH$, and $H_{2}O$, demonstrated enhanced precision and reduced computational overhead, key considerations given the limitations of current quantum hardware.

Extending these ideas, Sureshbabu \textit{et al.} \cite{Sureshbabu} applied a hybrid Quantum ML approach to simulate two-dimensional crystal structures, including monolayer hexagonal boron nitride ($h-BN$) and Graphene ($h-C$). Using a combination of RBMs and quantum circuits on IBM's quantum processors, they achieved results that closely mirrored those obtained through classical computational chemistry methods.

As these techniques evolved, researchers also turned their focus to optimizing VQE ansatz design and its compatibility with quantum hardware. Naeij \textit{et al.} \cite{Naeij} adopted the Unitary Coupled Cluster for Single and Double excitations (UCCSD) ansatz and implemented it through fermion-to-qubit transformations, applying it to molecules like $H_{3}^+$, $OH^-$, $HF$, and $BH_{3}$. Their benchmarking against classical methods such as Unrestricted HF (UHF) and FCI confirmed the potential of VQE to yield high-fidelity results.

Beyond ansatz optimization, new embedding and partitioning strategies have emerged to enhance scalability. Rossmannek \textit{et al.} \cite{Rossmannek} proposed a novel approach in which a mean-field potential was incorporated within a restricted action space, allowing them to define an effective Hamiltonian $\hat{H}_0$ whose ground state could be determined using VQE. Meanwhile, Song \textit{et al.} \cite{Song} demonstrated the practical utility of plane-wave basis sets in VQE simulations by replicating correlation optimised virtual orbital results within an $11 Ha (\approx 0.2993 eV)$ tolerance, using the Quantinuum H1-1 ion-trap quantum computer. These examples underscored the capacity of hardware-specific approaches to extend the applicability of VQE.
\begin{figure*}[b]
    \centering
    \includegraphics[width=\linewidth]{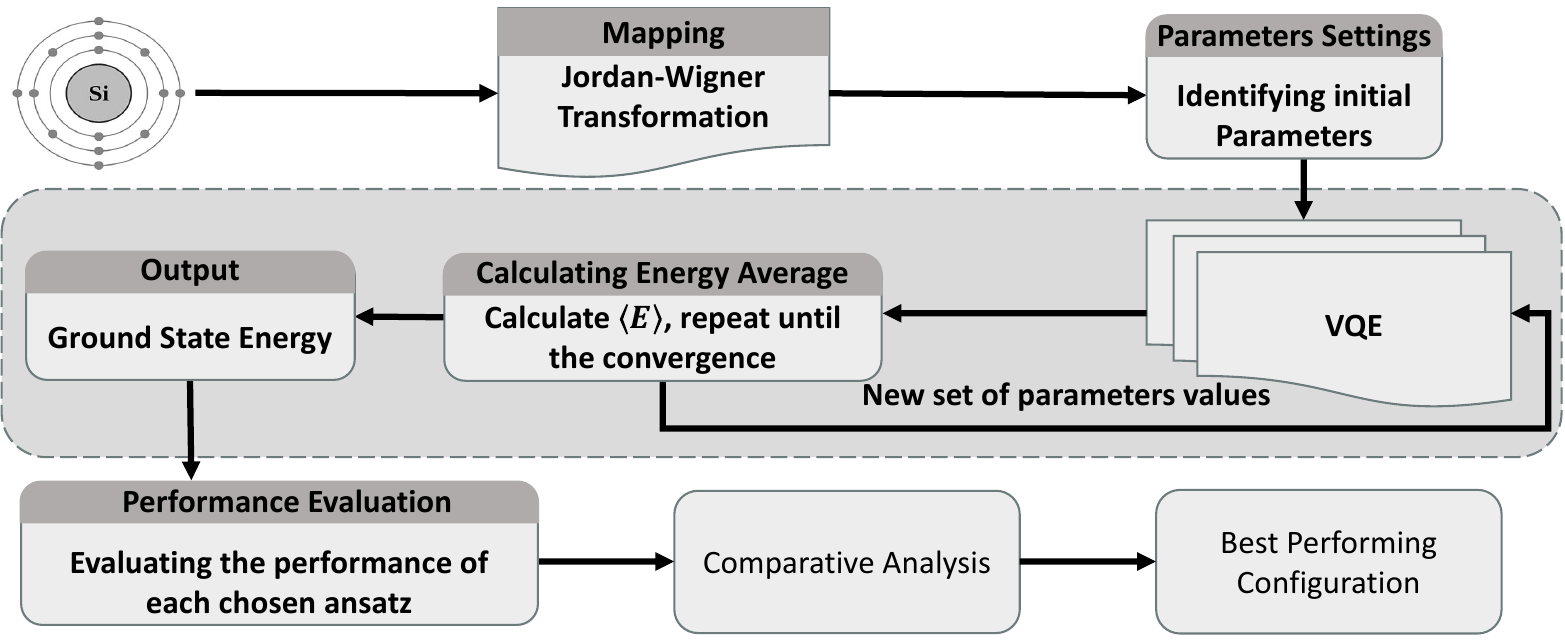}
    \caption{An overview of our proposed methodology.}
    \label{methodology}
\end{figure*}
Further refining VQE's performance, Han \textit{et al.} \cite{ZengH} proposed an optimized cost function within the meta-VQE framework. By introducing additional precision indicators and integrating quantum architecture search, they enhanced the method’s ability to learn ground-state energy behavior even in the presence of noise, a particularly relevant feature for NISQ devices. In another practical demonstration, Jerimiah \textit{et al.} \cite{Wright} investigated the role of gate-induced noise, ansatz circuit depth, and parameter optimization on VQE performance for sodium hydride. Their results provided valuable insight into designing ansatz circuits that maintain fidelity under noisy conditions.

Collectively, these studies show how quantum chemistry simulations have progressed from early QPE demonstrations to VQE approaches refined through ML, improved ansatz designs, measurement strategies, and noise mitigation. Applying these ideas to medium-sized systems, such as silicon, requires careful consideration of how circuit design, parameter initialization, and optimization methods work together to obtain reliable ground-state energy results.

\section{Methodology \label{sec3}}

We employ the VQE to estimate the ground-state energy of silicon-based molecular systems under different configurations. As shown in Fig.~\ref{methodology}, the overall workflow comprises several interconnected stages. The process begins with the construction of the molecular Hamiltonian, which is then mapped onto qubits to enable quantum representation. A parameterized quantum circuit is designed to model the system’s wavefunction, and its parameters are iteratively optimized within a hybrid quantum–classical loop to minimize the system’s expected energy. The resulting energy estimates are subsequently benchmarked against classical reference values to assess both precision and computational efficiency. The following subsections elaborate on each stage in detail, covering system preparation and Hamiltonian formulation, mapping techniques, ansatz architectures, parameter initialization strategies, optimization algorithms, and performance evaluation metrics.

\subsection{Hamiltonian Construction}
The electronic structure of the silicon-based system is retrieved using atomic and molecular data from the PubChem database. To simplify the quantum simulation, the Born–Oppenheimer approximation is employed, assuming that the nuclei remain fixed due to their much larger mass compared to electrons. The resulting electronic Hamiltonian is expressed in its second-quantized form as:
\begin{equation}
    \hat{H} = \sum_{p,q} h_{pq} a_p^\dagger a_q 
    + \tfrac{1}{2} \sum_{p,q,r,s} h_{pqrs} a_p^\dagger a_q^\dagger a_r a_s ,
\end{equation}
where $a_p^\dagger$ and $a_q$ are fermionic creation and annihilation operators, and $h_{pq}, h_{pqrs}$ represent one- and two-electron integrals (e.g., STO-3G basis), incorporating kinetic energy, electron–nucleus attraction, and electron–electron repulsion.

\subsection{Fermion-to-Qubit Mapping}
To simulate this fermionic Hamiltonian on a quantum computer, it is transformed into an operator acting on qubits using the Jordan–Wigner transformation:
\begin{equation}
a_i = \left( \prod_{j=0}^{i-1} Z_j \right) \cdot \frac{X_i + iY_i}{2}, \quad 
a_i^\dagger = \left( \prod_{j=0}^{i-1} Z_j \right) \cdot \frac{X_i - iY_i}{2},
\end{equation}
where $X_i, Y_i, Z_i$ are the Pauli matrices acting on qubit $i$. Applying this mapping to all fermionic terms yields a qubit Hamiltonian expressed as a weighted sum of Pauli strings:
\begin{equation}
    \hat{H} = \sum_k c_k P_k, \quad P_k \in \{I, X, Y, Z\}^{\otimes n},
\end{equation}
where $P_k$ denotes an $n$-qubit Pauli string composed of tensor products of the single-qubit operators ${I, X, Y, Z}$, and $c_k$ are real coefficients determined by the underlying molecular integrals. These coefficients encode the physical interactions of the electronic system after fermion-to-qubit mapping. The resulting qubit Hamiltonian serves as the cost function in the VQE.
\begin{figure*}[htpb]
 \centering
 \includegraphics[width=\linewidth]{ 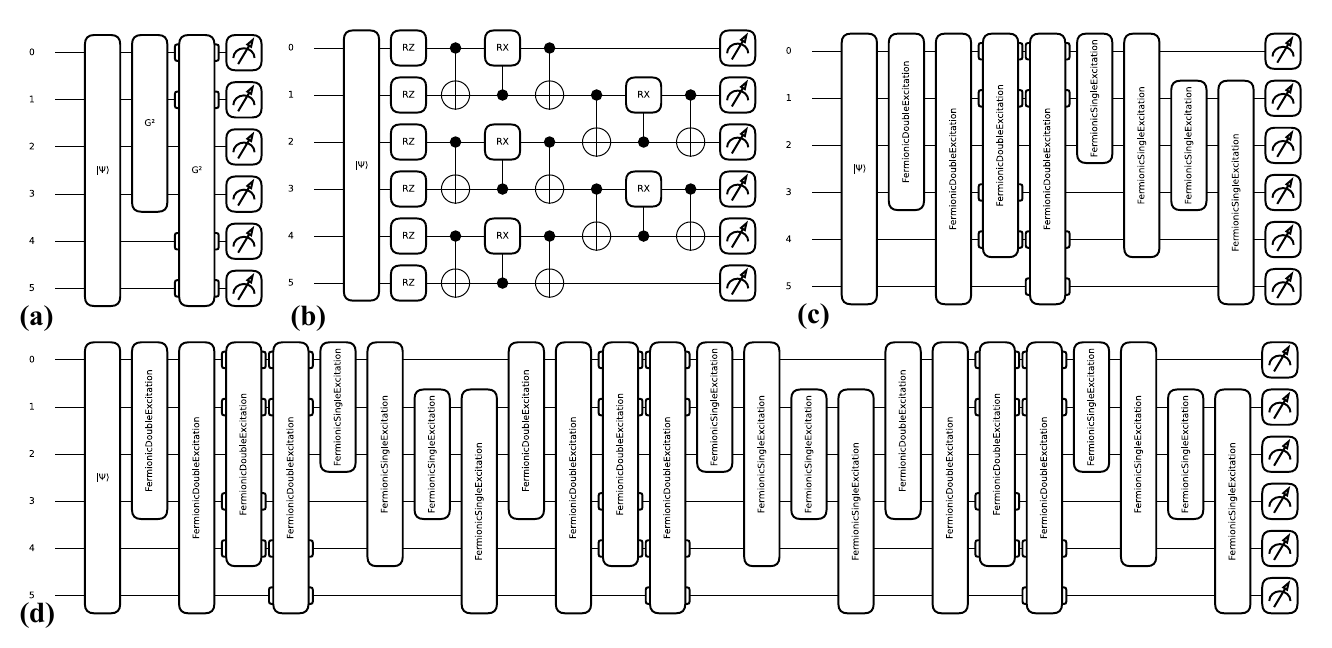}
 \caption{Circuit structures of the ansatz architectures investigated in this study: (a) Double Excitation Gates Ansatz, (b) Particle-Conserving U2 Ansatz (composed of RX, RZ, and CNOT gates), (c) UCCSD Ansatz (including Fermionic Single and Double excitation gates), and (d) k-UpCCGSD Ansatz (implemented with three layers of the UCCSD ansatz). }
 \label{quantum_circuits}
 \end{figure*}
 
\subsection{Variational Ansatz}

The trial wavefunction in VQE is represented by a parameterized quantum circuit, or ansatz, which defines the variational search space. The choice of ansatz is crucial, as it directly influences the precision, trainability, and feasibility of the simulation. In near-term quantum computing, a central challenge lies in balancing physical precision with hardware limitations.
Broadly, ansatz architectures can be categorized into two classes. Hardware-Efficient ansatz, such as ParticleConservingU2, are designed to be shallow and implementable on NISQ devices. Their reduced depth and fewer parameters make them more resilient to noise, although they often lack chemical intuition and can suffer from barren plateaus during optimization. By contrast, chemically inspired ansatz, such as UCCSD and k-UpCCGSD, are derived from many-body theory and can accurately model electron correlation. However, they require deeper circuits, which makes them more susceptible to decoherence and gate errors. Positioned between these extremes, the DoubleExcitation Gates (DexcG) ansatz offers a hybrid compromise, combining chemical relevance with shallow depth.

In this work, we compare four representative ansatz architectures (see Fig.~\ref{quantum_circuits}):
\begin{itemize}
 \item \textbf{DoubleExcitation Gates (DexcG)} \cite{Loos}: A low-depth ansatz incorporating only two-electron excitation gates. It is straightforward to implement and robust against noise, but its limited expressibility restricts its ability to capture strong correlation effects.

\item \textbf{ParticleConservingU2}:  A hardware-efficient, symmetry-preserving ansatz that conserves particle number. Its structure reduces the variational search space and improves trainability, though barren plateaus can still hinder convergence.  

\item \textbf{UCCSD (Unitary Coupled Cluster Singles and Doubles)} \cite{Xia}:   A chemically motivated ansatz that accurately models correlated wavefunctions. The trial state is given by  
\begin{equation}  
   |\psi_{UCCSD}\rangle = e^{T - T^\dagger} | \psi_{HF} \rangle ,  
\end{equation}  
where \( |\psi_{HF}\rangle \) is the Hartree–Fock reference state, and \(T\) is the cluster operator containing single (\(T_1\)) and double (\(T_2\)) excitation terms. The Hermitian conjugate \(T^\dagger\) ensures unitarity of the operator. While accurate, UCCSD circuits are deep and often exceed the coherence limits of current hardware.  

\item \textbf{k-UpCCGSD (k-fold Unitary Pair Coupled Cluster with Generalized Singles and Doubles)} \cite{Lee J}:  
An extension of UCCSD that applies generalized excitation operators repeatedly \(k\) times:  
\begin{equation}  
    |\psi_{k\text{-UpCCGSD}}\rangle = \prod_{i=1}^k e^{T - T^\dagger} | \psi_{HF} \rangle ,  
\end{equation}  
where \(k \in \mathbb{Z}^+\) is a tunable repetition factor controlling the circuit depth and expressibility. The generalized cluster operator \(T\) includes both particle-conserving single and pair excitations. By adjusting \(k\), one can balance precision against hardware resource requirements.  
\end{itemize}

\subsection{Parameter Initialization}
The initialization of circuit parameters is an important design choice in the VQE algorithm, as it influences the optimization trajectory and the exploration of the cost-function landscape. Different initialization strategies may lead to different convergence behaviors, especially for expressible ansatz architectures with complex, non-convex energy surfaces. To investigate these effects, we consider four schemes:
\begin{enumerate}
 \item \textbf{All-zero initialization}: every variational parameter is set to zero at the start. 
\item \textbf{All-half initialization}: all parameters are initialized to the fixed value \(0.5\), placing the circuit in a balanced configuration between trivial and fully rotated parameter values.  
\item \textbf{All-one initialization}: each parameter is initialized with the value \(1\), ensuring a non-trivial starting point distinct from the zero case.  
\item \textbf{Random initialization}: parameters are sampled independently from a uniform distribution, \(\theta_i \sim \mathcal{U}(0,1)\). This introduces stochastic variability between runs and provides insight into the role of randomness in convergence stability.  
\end{enumerate}

\subsection{Optimization}

Optimization is a fundamental component of the VQE, serving as the classical engine that iteratively updates the parameters of the quantum circuit in order to minimize the expectation value of the Hamiltonian. This minimization directly corresponds to approximating the ground state energy of a quantum system. Formally, the optimization seeks to solve:
\begin{equation}
    \min_{\theta} E(\theta) = \langle \psi(\theta) | \hat{H} | \psi(\theta) \rangle, 
\end{equation}
where $\ket{\psi(\theta)}$ is the variational quantum state prepared by the parameterized ansatz, $\theta$ is a real-valued parameter vector, and $\hat{H}$ is the Hamiltonian. Due to the high-dimensional, non-convex, and typically noisy landscapes of VQE cost functions, particularly with expressible ansatz architectures such as UCCSD or k-UpCCGSD, the choice of the classical optimizer plays a decisive role in the convergence behavior and final energy estimation.

In our methodology, we contrast three widely used optimization procedures: Gradient Descent (GD), Simultaneous Perturbation Stochastic Approximation (SPSA), and Adaptive Moment Estimation (ADAM). Each of the three optimizers has advantages and limitations within the quantum simulation framework.
\subsubsection{Gradient Descent}

GD \cite{Ruder} is a deterministic, first-order optimization technique that updates parameters in the direction of the steepest descent. It is defined by the update rule:
\begin{equation}
   \theta_{t+1} = \theta_t - \eta \nabla E(\theta_t), 
\end{equation}
where $\eta$ is the learning rate, and $\nabla E(\theta_t)$ is the gradient of the cost function with respect to the parameters at iteration $t$. 
Its simplicity and theoretical convergence guarantees make it attractive for smooth and well-behaved cost landscapes. However, in the VQE context, GD is often hindered by challenges such as vanishing gradients, sensitivity to local minima, and noise sensitivity, especially when highly expressible ansatz architectures are employed.

\subsubsection{Simultaneous Perturbation Stochastic Approximation}

SPSA \cite{Spall} is a gradient-free stochastic optimization method particularly well-suited for high-dimensional, noisy objective functions. Instead of computing the full gradient, SPSA approximates it using only two evaluations per iteration, regardless of the number of parameters. The update rule is given by:
\begin{equation}
    \theta_{t+1} = \theta_t - \eta \cdot \hat{g}_t,
\end{equation}
where the gradient approximation $\hat{g}_t$ is:
\begin{equation}
   \hat{g}_t = \frac{E(\theta_t + c_t \Delta_t) - E(\theta_t - c_t \Delta_t)}{2 c_t} \cdot \Delta_t^{-1},
\end{equation}
 with $c_t$ a small perturbation parameter and $\Delta_t$ a random vector with independent symmetric components (e.g., $\pm 1$). 
Its robustness to measurement noise and stochastic behavior makes it particularly effective when optimizing expressible circuits, where traditional gradient-based methods struggle.

\subsubsection{Adaptive Moment Estimation}

ADAM \cite{Kingma} is a stochastic optimization algorithm that combines the strengths of Momentum and RMSProp \cite{Tieleman}, by adjusting parameter updates using first and second moment estimates of the gradients, thus stabilizing and accelerating convergence. 
The parameter update rule is given by:
\begin{equation}
    \theta_{t+1} = \theta_t - \eta \cdot \frac{\hat{m}_t}{\sqrt{\hat{v}_t} + \epsilon},
\end{equation}
where $\hat{m}_t$ and $\hat{v}_t$ are bias-corrected first and second moment estimates of the gradient. ADAM is advantageous in handling irregular landscapes and adapting learning rates, which is particularly beneficial when optimizing deep quantum circuits in noisy environments, as is often the case in VQE.


\subsection{Ground-State Estimation and Performance Evaluation}

After the VQE loop converges, the optimized parameters yield an estimate of the molecular ground-state energy $\hat{E}$. To assess the quality of these results, we focus on two aspects: energy error and convergence.

The energy error is defined as the deviation between the estimated energy and experimentally reported values $E_{\mathrm{exp}}$. We use the relative error, 

\begin{equation}    
\Delta E_{\mathrm{rel}} = \frac{\lvert \hat{E} - E_{\mathrm{exp}} \rvert}{\lvert E_{\mathrm{exp}} \rvert} ,
\end{equation}

as the primary metric for quantifying the physical reliability of the simulation.

Convergence is examined in terms of the number of iterations required to reach stability, the smoothness of the optimization trajectory, and the variability of the final energy values. This allows us to capture not only whether the algorithm reaches the correct solution, but also how efficiently and consistently it does so.

The analysis is carried out across different parameter initialization strategies (zeros, ones, half-values, random), ansatz architectures (DexcG, ParticleConservingU2, UCCSD, k-UpCCGSD), and optimizers (GD, SPSA, ADAM). By comparing results along these three dimensions, we identify the configurations that yield the lowest energy errors with the most reliable convergence behavior for simulating silicon-based systems.

\section{Results and Discussion \label{sec4}}

\subsection{Experimental Settings}
Table~\ref{tab:exp-settings} summarizes the quantum chemistry specifications, VQE configurations, and execution details used in our simulations of the silicon atom. These settings provide the full reproducibility context for the reported results.  
\begin{table}[!h]
\centering
\caption{Simulation setup and hyperparameters for the VQE experiments on the silicon atom.}
\label{tab:exp-settings}
\begin{tabular}{p{0.38\linewidth} p{0.55\linewidth}}
\toprule
\textbf{Parameter} & \textbf{Value / Setting} \\
\midrule
Molecule & Silicon (Si), details from PubChem \cite{Kim} \\
Electronic Structure Method & Hartree–Fock (reference), Experimental data (benchmark) \cite{NIST, CODATA}\\
Basis Set & STO-3G (minimal basis) \\
Number of Qubits & $18$ qubits (Jordan–Wigner mapping, no qubit reduction) \\
\midrule
Ansatz & DexcG, PCU2, UCCSD, k-UpCCGSD \\
Optimizer & GD, SPSA, ADAM \\
Learning Rate & $0.5$ (for gradient-based optimizers)\\
Number of Iterations & $50$ (per optimization run) \\
Initialization & $\{0,\,0.5,\,1,\,100\}$ (different seeds) \\
\midrule
Measurement Scheme & Expectation values via Pauli-term sampling \\
Number of Shots & 1024 (per expectation value) \\
Software Framework & \texttt{PennyLane} \cite{Bergholm} \\
Quantum Backend & \texttt{lightning.qubit} (state-vector simulator) \\
Classical Hardware & Kaggle Cloud environment: Intel Xeon CPU (2 vCPUs), $13\sim30$ GB RAM \\
\bottomrule
\end{tabular}
\end{table}

\subsection{Parameter Initialization Analysis}
\begin{figure*}[htpb]
    \centering
    \includegraphics[width=\linewidth]{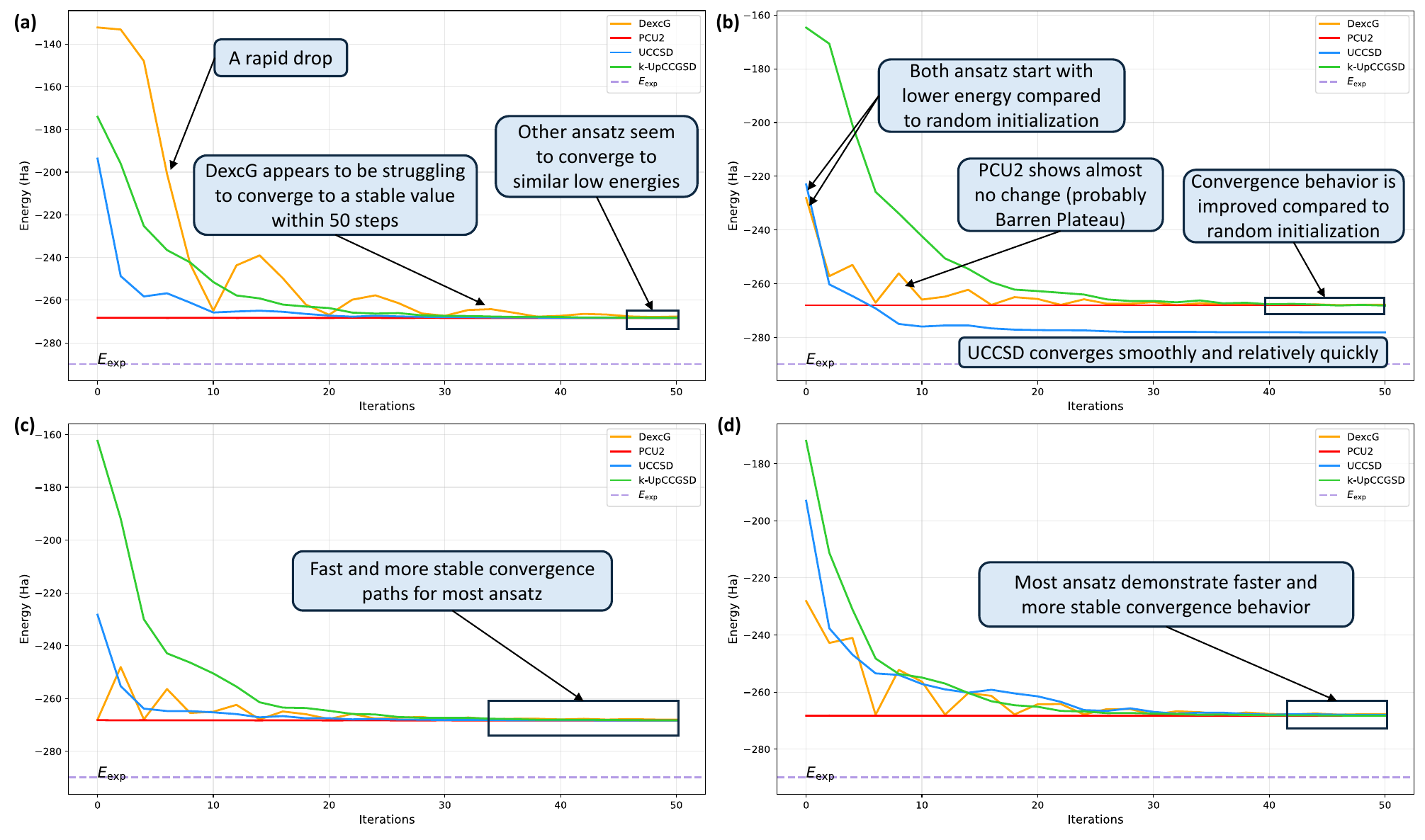}
    \caption{Energy convergence comparison for different parameter initialization strategies across the selctect ansatz: (a) Random, (b) Zero, (c) Half, and (d) One.}
    \label{init}
\end{figure*}

In this set of experiments, we address the question: \textit{How do different initialization strategies affect convergence across the selected ansatz architectures?} To answer this, the optimizer is fixed to ADAM, and convergence is compared under four initialization schemes: random, zero, half, and one (Fig. \ref{init}). The results are organized such that each initialization is shown collectively for all four ansatz architectures, allowing a direct assessment of which strategies are most effective across circuit types.
When all variational parameters are randomly initialized (Fig. \ref{init}-a), convergence paths are unstable, especially for the DexcG and k-UpCCGSD architectures, which exhibit substantial oscillations and slower stabilization. DexcG, in particular, fails to reach a stable energy even after 50 optimization steps, suggesting a complex energy landscape with sensitivity to local minima or saddle points.

In contrast, UCCSD and PCU2 show smoother and more stable convergence under random initialization, with PCU2 reaching low-energy values rapidly. When the initial parameters are set deterministically to $0$ (Fig. \ref{init}-b), $0.5$ (Fig. \ref{init}-c), or $1$ (Fig. \ref{init}-d), all ansatz architectures, especially UCCSD, exhibit significantly improved convergence behavior, most notably under zero initialization. PCU2 remains highly stable, maintaining a nearly constant low energy across all fixed initializations, implying a relatively smooth optimization landscape or low sensitivity of the cost function to parameter variations. UCCSD and k-UpCCGSD also benefit from deterministic initialization, achieving rapid and smooth convergence toward consistent final energies that are nearly indistinguishable across the fixed-value scenarios. Although DexcG shows improvement under deterministic initialization relative to its random counterpart, it remains the most oscillatory and slowest to converge among the tested ansatz architectures, reinforcing its sensitivity to initial conditions and indicating a non-convex optimization landscape with multiple local minima.

Importantly, across non-random initializations, the final converged energies of PCU2, UCCSD, and k-UpCCGSD are comparable and approach the experimental reference value, confirming the effectiveness of these ansatz architectures and the optimizer in capturing the ground-state energy.
These findings collectively emphasize that careful selection of initial parameters significantly enhances the stability, speed, and reliability of VQE optimizations. The study thus underscores the broader necessity of initialization strategies tailored to the ansatz architecture in quantum algorithm design, particularly in quantum chemistry applications where convergence to the actual ground state is crucial.

\subsection{Ansatz Choice Analysis}
\begin{figure*}[htpb]
    \centering
\includegraphics[width=1\linewidth]{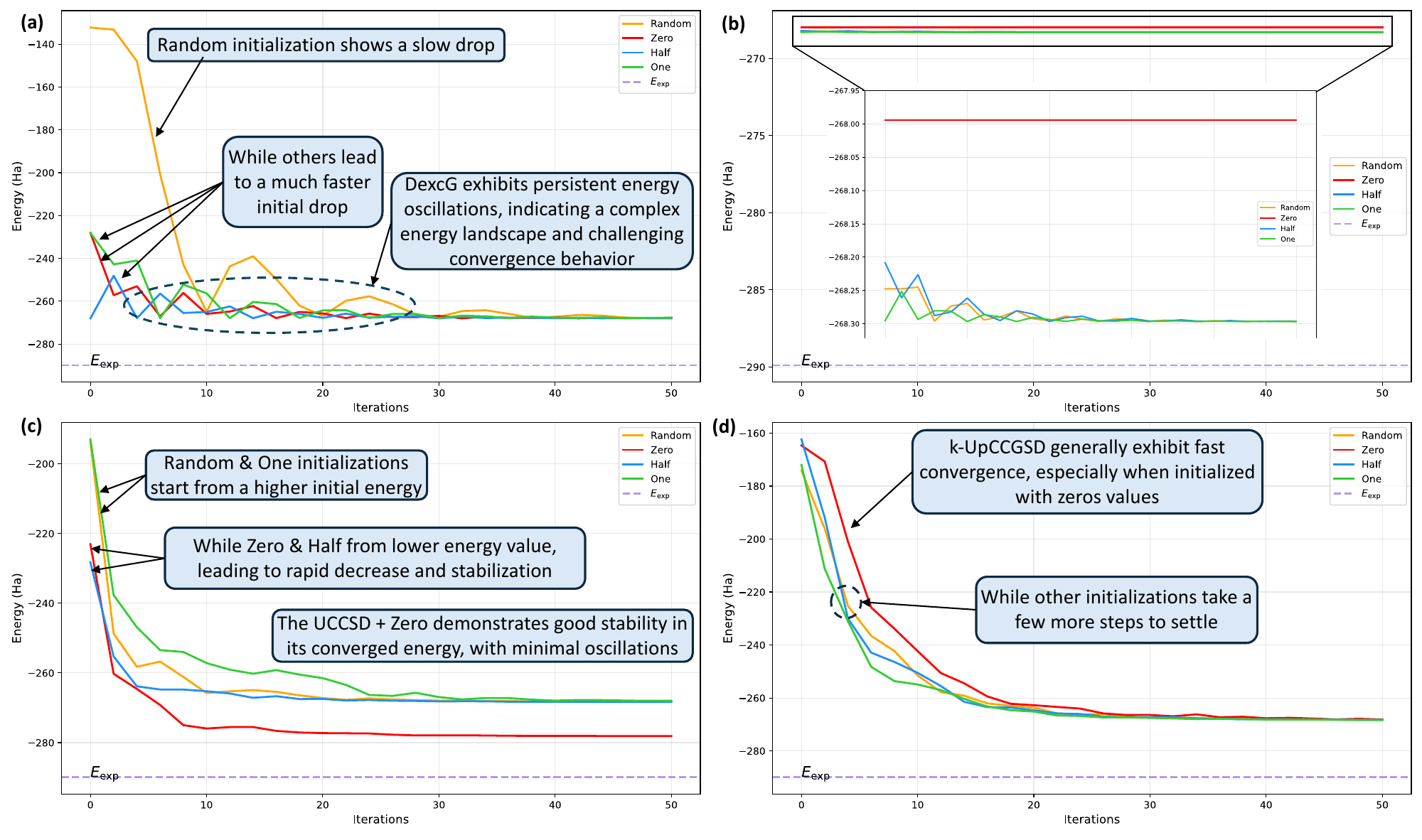}
    \caption{Energy convergence comparison across different ansätze with varying parameter initialization: (a) Double Excitation Gates, (b) ParticleConservingU2, (c) UCCSD, and (d) k-UpCCGSD.The results highlight the impact of ansatz structure on convergence behavior and optimization stability. }
    \label{ansatz}
\end{figure*}

In this set of experiments, we ask the complementary question: \textit{For a given ansatz architecture, how sensitive is convergence to the choice of initialization?} Using the same ADAM optimizer, the results are reorganized so that each ansatz architecture is shown with its four initialization strategies (Fig. \ref{ansatz}). This representation emphasizes the robustness, or fragility, of individual ansatz architectures to parameter initialization, in contrast to the previous subsection, which focuses on comparing initialization schemes across all architectures.

Random initialization consistently leads to unstable and oscillatory convergence, most notably for the DexcG architecture, which exhibits large fluctuations and a delayed descent in energy (Fig. \ref{ansatz}-a). This behavior illustrates its high sensitivity to initial conditions.
By contrast, deterministic initializations ($0$, $0.5$, or $1$) markedly improve convergence speed and stability across all ansatz architectures. PCU2 stands out as uniquely robust, displaying consistently low and stable energy profiles regardless of initialization (Fig. \ref{ansatz}-b). Both UCCSD and k-UpCCGSD also respond favorably to deterministic initializations, particularly zero initialization, where they achieve rapid and smooth convergence to low-energy states indicative of precise ground-state approximations (Fig. \ref{ansatz}-c, \ref{ansatz}-d). Although DexcG improves under deterministic initialization, it remains the most fragile architecture, converging more slowly and less reliably than the others.

Taken together, these results show that ansatz architecture plays a decisive role in determining robustness to initialization. PCU2 is consistently stable and initialization-independent, UCCSD benefits strongly from appropriate initialization (especially zero initialization), and DexcG highlights the difficulties posed by architectures with complex optimization landscapes. This perspective complements the previous subsection by shifting the focus from ``which initialization strategy works best overall'' to ``which ansatz architectures are most robust to initialization''. 

\subsection{Optimizer Analysis}
\begin{figure*}[htpb]
    \centering
    \includegraphics[width=\linewidth]{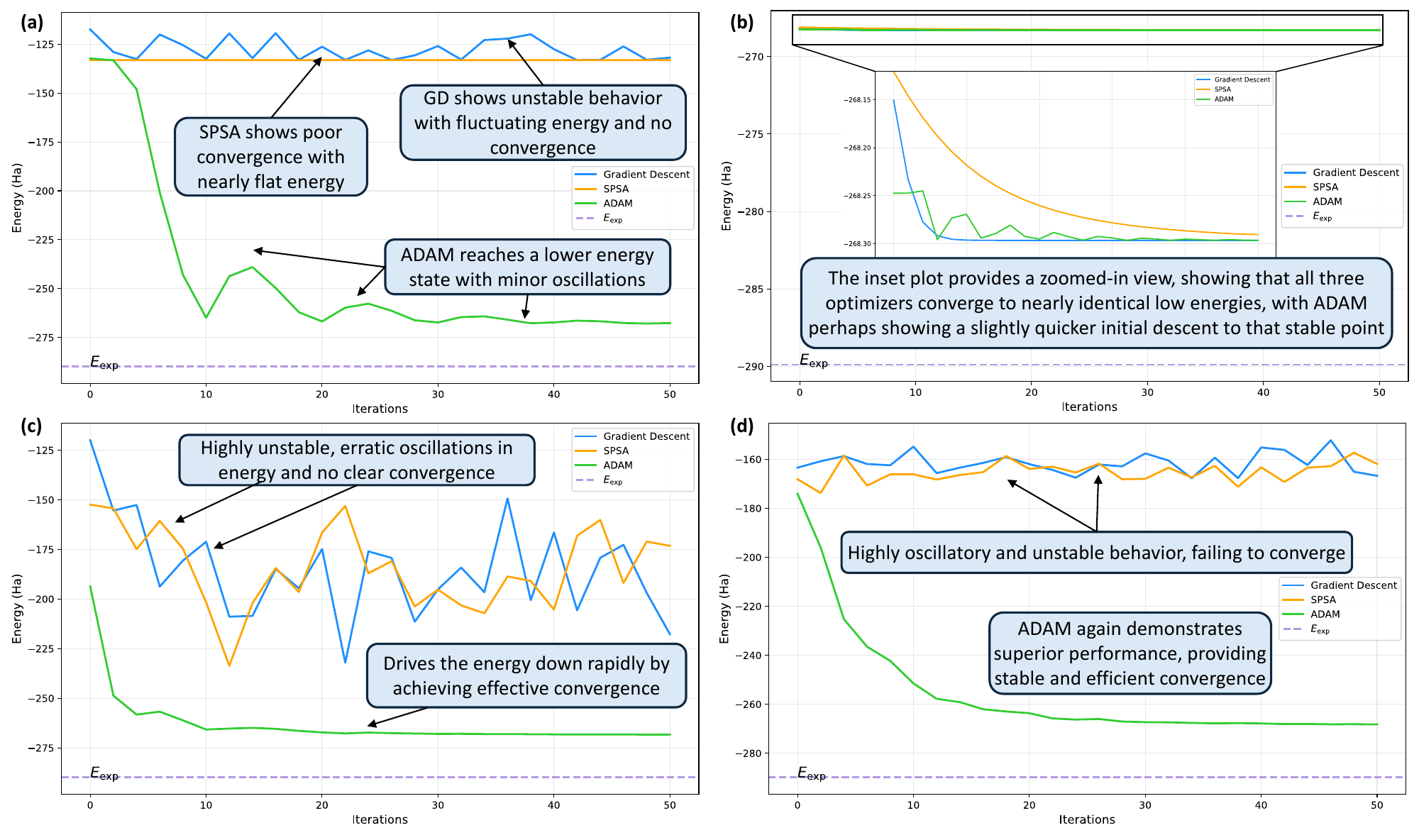}
    \caption{Energy convergence comparison across different ansätze using various optimizers: (a) Double Excitation Gates, (b) ParticleConservingU2, (c) UCCSD, and (d) k-UpCCGSD. The analysis reveals how the choice of optimizer interacts with the ansatz structure to influence convergence behavior and final energy precision.}
    \label{opt}
\end{figure*}

In this set of experiments, we address the following question: \textit{How does the choice of optimizer influence convergence behavior?} To isolate this effect, the initialization is fixed to the random scheme, and GD, SPSA, and ADAM are compared across the four ansatz architectures (Fig. \ref{opt}). This analysis highlights how optimizer dynamics interact with circuit complexity to determine stability and convergence speed.
For DexcG, UCCSD, and k-UpCCGSD, both gradient descent and SPSA exhibit unstable convergence in Fig. \ref{opt}-a, \ref{opt}-c, and \ref{opt}-d, characterized by noticeable oscillations and an inability to reach low-energy minima within 50 optimization steps.
In contrast, ADAM consistently reduces the energy to lower and more stable values, leveraging its adaptive learning rates ($\theta_{t+1}$) and momentum terms ($m_{t}, v_{t}$) to effectively navigate complex optimization landscapes.
Fig. \ref{opt}-b shows PCU2 as a notable exception, where all three optimizers converge efficiently to nearly identical low-energy states, suggesting a smoother and more favorable energy landscape. Overall, ADAM emerges as the most robust and effective optimizer, particularly in scenarios involving challenging variational landscapes or randomized initial conditions, reinforcing its importance for achieving precise and stable quantum simulations in chemistry and materials science.

\subsection{Discussion}

This study highlights the intricate interplay between parameter initialization, ansatz architecture, and optimizer selection, all of which critically shape the performance of the VQE in simulating the ground-state energy of complex molecular systems such as the silicon atom.

Our results clearly demonstrate that the parameter initialization strategy significantly affects both the convergence behavior and the precision of the final energy. In particular, random initialization often leads to unstable or oscillatory optimization trajectories, especially for complex ansatz like DexcG and k-UpCCGSD, which appear to operate within rugged optimization landscapes. These erratic patterns reflect the sensitivity to local minima and saddle points, common obstacles in high-dimensional variational circuits.

In contrast, deterministic initializations (i.e., setting all parameters to fixed values such as 0, 0.5, or 1) that profoundly improve convergence performance. Among these, zero initialization emerges as the most robust and generally effective scheme. 
This is especially the case for the UCCSD ansatz, which achieves smooth, stable, and rapid convergence to low-energy states, even in the presence of complex parameter dependencies. In particular, the PCU2 ansatz demonstrates exceptional consistency across all tested initialization parameters, suggesting that its energy landscape is relatively smooth and less sensitive to the starting point of optimization. Although DexcG showed moderate improvements under fixed initialization, it remained the most challenging ansatz to reliably optimize, demonstrating its limited applicability without tailored optimization strategies.

Beyond initialization, our comparative analysis of optimizers reveals that ADAM offers significant advantages over traditional approaches such as gradient descent and SPSA. ADAM’s adaptive learning rate and momentum accumulation capabilities enable it to navigate through complex, non-convex landscapes efficiently, resulting in faster and more stable convergence across a broad range of ansatz. This effect is most pronounced for DexcG and k-UpCCGSD, where common optimizers struggled to reach low-energy solutions within the same number of iterations. Interestingly, PCU2 once again stands out as an exception, converging effectively across all optimizers, another indication of its favorable landscape characteristics.

Together, these results strongly support the conclusion that co-designing the initialization scheme, ansatz architecture, and optimizer is essential for achieving high-convergence and efficient quantum simulations. The configuration combining zero initialization, the UCCSD ansatz, and the ADAM optimizer consistently yields the most stable and precise ground-state energies, closely approximating experimental reference values with reduced optimization overhead.

These findings reinforce the broad theme of this work: To unlock the full potential of VQE for real-world physics and material science applications, especially in the near-term noisy intermediate-scale quantum (NISQ) era, one must go beyond generic algorithmic choices and adopt tailored, synergistic strategies. With advancing quantum hardware, such methodology-guided tuning will be essential to make scalable, chemistry-reliable simulations of intractable systems available for classical algorithms.

\begin{table}[htbp]
\centering
\caption{Relative Error (\%) for VQE with different initializations, ansatz, and optimizers}
\label{tab:re_results}
\begin{tabular}{@{}llllll@{}}
\toprule
\textbf{Initialization} & \textbf{Ansatz} & \textbf{GD} & \textbf{SPSA} & \textbf{ADAM} \\
\midrule
Random & DexcG & 54.52 & 54.08 & 7.67 \\
       & PCU2  & 7.45  & 7.45  & 7.45 \\
       & UCCSD & 24.91 & 40.25 & 7.45 \\
       & k-UpCCGSD & 42.48 & 44.16 & 7.47 \\
\midrule
Zero   & DexcG & 57.86 & 54.08 & 7.62 \\
       & PCU2  & 7.55  & 7.55  & 7.55 \\
       & UCCSD & 12.16 & 17.69 & 4.06 \\
       & k-UpCCGSD & 42.03 & 44.39 & 7.50 \\
\midrule
Half   & DexcG & 54.08 & 54.08 & 7.56 \\
       & PCU2  & 7.45  & 7.45  & 7.45 \\
       & UCCSD & 23.11 & 30.60 & 7.44 \\
       & k-UpCCGSD & 43.19 & 41.72 & 7.45 \\
\midrule
One    & DexcG & 58.35 & 54.08 & 7.65 \\
       & PCU2  & 7.45  & 7.45  & 7.45 \\
       & UCCSD & 38.23 & 34.81 & 7.55 \\
       & k-UpCCGSD & 45.89 & 40.83 & 7.45 \\
\bottomrule
\end{tabular}
\label{RE}
\end{table}

From the graphs and Table \ref{RE}, which summarize the impact of the previously discussed factors through relative error (RE) on the ground-state energy of the Silicon atom, the PCU2 ansatz was the most robust, achieving $\sim 7.45$\% RE across all optimizers and starting values, and both expressed well and having a good optimization landscape. While, DexcG was consistently bad with errors $>54\%$, indicating that it had too little expressibility. The k-UpCCGSD ansatz worked competitively only in conjunction with the ADAM optimizer ($\sim 7.45–7.50\%$ RE) but fared poorly under GD and SPSA, which pointed to its dependency upon adaptive momentum-based algorithms. The UCCSD ansatz also showed the highest variance: though its best configuration (Zero initialization with ADAM) recorded the minimum error ($4.06\%$ RE), it tended to converge to non-optimal minima using different configurations, which further emphasized its sensitivity to initial conditions and the choice of optimizer.

\subsection{Key Summary Results}
Based on the experimental results, the principal findings of this study can be summarized as follows:

\begin{itemize}
    \item A joint co-design of the initialization scheme, ansatz architecture, and optimization algorithm is essential to achieve stable and efficient quantum simulations.
    \item The configuration combining zero-initialization, the UCCSD ansatz, and the ADAM optimizer provides the most accurate and consistent ground-state energy estimates.
    \item The initialization strategy exerts a significant influence on both convergence behavior and energy precision.
    \item Zero initialization facilitates smooth and rapid convergence, particularly when used with the UCCSD ansatz.
    \item Random initialization often leads to unstable or oscillatory optimization trajectories, impeding convergence.
    \item The PCU2 ansatz demonstrates high consistency and low sensitivity to initialization variations.
    \item The DexcG ansatz remains challenging to optimize without problem-specific or tailored strategies.
    \item Among the tested optimizers, ADAM consistently outperforms gradient descent and SPSA in terms of both convergence rate and solution precision.
\end{itemize}

\section{Conclusion \label{sec6}}

This study provides a comprehensive evaluation of the VQE for quantum chemistry applications, with a particular focus on simulating challenging heavy-element systems such as silicon. We systematically assess the impact of various parameter initialization schemes, including random values and fixed initializations at $0, 0.5$, and $1$ in conjunction with a comparative analysis of four ansatz (Double Excitation Gates ``DexcG'', ParticleConservingU2 ``PCU2'', UCCSD, and k-UpCCGSD) and three prominent optimizers (gradient descent, SPSA, and ADAM). Our findings reveal critical interdependencies between parameter initialization strategies, ansatz architecture, and optimizer behavior, all of which significantly influence the convergence efficiency and final precision of the VQE algorithm.

A key finding is the significant influence of initial parameter selection on optimization efficiency. Among all tested initialization schemes, zero initialization emerges as the most effective, consistently enabling smoother convergence and lower energy states across all ansatz. This effect is particularly pronounced for the UCCSD ansatz, which, when paired with zero initialization, demonstrates remarkable stability and convergence performance.

The optimizer analysis further reinforces the superiority of the ADAM optimizer, which consistently outperforms both gradient descent and SPSA, especially in navigating rugged and highly non-convex energy landscapes. When taken together, the optimal configuration for VQE in this study is the combination of “Zero Initialization, UCCSD Ansatz, ADAM Optimizer”, which delivers the most precise and stable results with minimal optimization overhead.

These results underscore the importance of co-designing ansatz, optimizers, and initialization schemes for variational quantum algorithms. As quantum hardware continues to evolve toward greater scalability and fidelity, such careful orchestration will be essential for realizing a quantum advantage in real-world quantum chemistry problems. Ultimately, this work reinforces VQE’s promise as a viable and powerful approach for scalable molecular electronic structure calculations, capable of complementing and potentially surpassing classical methods in precision and efficiency.

 \section*{Acknowledgment}
 This work was carried out with the support of the National Center for Scientific and Technical Research (CNRST) as part of the ``PhD-Associate Scholarship – PASS'' Program.
 This work was also supported in part by the NYUAD Center for Quantum and Topological Systems (CQTS), funded by Tamkeen under the NYUAD Research Institute grant CG008.

\end{document}